\documentclass[paper]{elsarticle}

\usepackage{amsfonts,amsbsy,amssymb,amsmath}
\usepackage{lineno,hyperref}
\usepackage{graphicx}
\usepackage[table,xcdraw]{xcolor}
\usepackage{float}

\graphicspath{{figures/}}

\modulolinenumbers[5]

\journal{Chemical Physics Letters}

\biboptions{compress}

\begin{document}

\begin{frontmatter}

\title{The Phase Space Mechanism for Selectivity \\ in a Symmetric Potential Energy Surface with a Post-Transition-State Bifurcation}

 \author[label1]{M. Agaoglou}

 \author[label2]{V. J. Garc\'ia-Garrido}
 \author[label1]{M. Katsanikas}
 \author[label1]{S. Wiggins\corref{mycorrespondingauthor}}
 \ead{S.Wiggins@bristol.ac.uk}

  \address[label1]{School of Mathematics, University of Bristol, \\ Fry Building, Woodland Road, Bristol, BS8 1UG, United Kingdom.\\[.2cm]}

 \address[label2]{Departamento de F\'isica y Matem\'aticas, Universidad de Alcal\'a, \\ Alcal\'a de Henares, 28871, Spain.}


\begin{abstract}
Chemical selectivity is a phenomenon displayed by potential energy surfaces (PES) that is relevant for many organic chemical reactions whose PES feature a valley-ridge inflection point (VRI) in the region between two sequential index-1 saddles. In this letter we describe the underlying dynamical phase space mechanism that qualitatively determines the product distributions resulting from bifurcating reaction pathways. We show that selectivity is a consequence of the heteroclinic and homoclinic connections established between the invariant ma\-nifolds of the families of unstable periodic orbits (UPOs) present in the system. The geometry of the homoclinic  and heteroclininc connections is determined using the technique of Lagrangian descriptors, a trajectory-based scalar technique with the capability of unveiling the geometrical template of phase space structures that characterizes transport.
\end{abstract}


\end{frontmatter}


\section{Introduction}

A major topic in the study of organic chemical reactions is that of providing a theoretical understanding of the underlying mechanisms that govern selectivity, i.e.  product distributions. A significant amount of effort (\cite{rehbein2011,ess2008,hare2017post})  on this problem has been focused on a potential energy surface (PES) having two sequential index-1 saddles with no intervening energy minimum. Rather, between the two  index-1 saddles is a valley ridge inflection (VRI) point.  One saddle is of higher energy than the other. The reaction is initiated when a trajectory crosses the region of the higher energy saddle and approaches the lower energy saddle (the ``entrance channel''). On either side of the lower energy saddle are two wells. The question of interest is which trajectory does the well enter (``product selectivity'')?

There is a belief that the VRI plays a role in selectivity. Certainly the VRI is an essential geometrical feature of the PES that arises in the configurations of saddles and wells that we have just described.
Mathematically, at a VRI point two conditions are met: the Gaussian curvature of the PES is zero, which implies that the Hessian matrix has a zero eigenvalue, and also the gradient of the potential is perpendicular to the eigenvector corresponding to the zero eigenvalue. Geometrically, this means that the landscape of the PES in the neighborhood of the VRI changes its shape from a valley to a ridge. VRI points are ubiquitous in the chemistry literature and have attracted the attention of both chemists and mathematicians in the past decades \cite{Valtazanos1986,quapp2004,birney2010}. 
In the vicinity of VRI points the intrinsic reaction coordinate bifurcates due to the shape of the PES, and this gives rise to a reaction mechanism known as a two-step-no-intermediate mechanism \cite{singleton2003}. 

In this letter we provide a fundamental explanation of how selectivity arises as a mechanism in phase space. A basic introduction to the background and techniques for the phase space approach to chemical reaction dynamics can be found in \cite{Agaoglou2019}. 
Our work extends the work of \cite{collins2013} where an analysis of the influence of VRIs on the dynamics and product distributions was carried out by studying an ensemble of trajectories initialized on the dividing surface of the unstable periodic orbit (UPO) associated to the high-energy index-1 saddle at the entrance channel of the PES. 
There are two main differences between the work of \cite{collins2013} and the work described in this letter. The two potential wells in the PES that we consider are symmetric (described below more precisely) and we explicitly describe the phase space structures governing selectivity in terms of the geometric structures arising from the UPOs that control access to the entrance channel and the two wells. For the symmetric PES the product distributions are known in advance, but this has the advantage of providing us with the insight to understand how the phase space structures control selectivity.

We show that the scalar-based trajectory diagnostic technique known as Lagrangian descriptors \cite{madrid2009,mancho2013lagrangian,lopesino2017}
can reveal the geometrical template of phase space structures that characterizes the different transport routes in phase space. In particular, we show that selectivity arises in the phase space of the system as the result of the heteroclinic and homoclinic connections established between the invariant stable and unstable manifolds of all the families of UPOs  that control access to the entrance channel and the two wells.

This letter is outlined as follows. In Section \ref{sec:model} we introduce the relevant landscape features of the PES that determines the two degree-of-freedom (DoF) Hamiltonian model used to understand the emergence of selectivity in these type of chemical systems. Section \ref{Tr_mech} is devoted to describing how the branching mechanism arises in phase space from the heteroclinic and homoclinic intersections between the stable and unstable manifolds associated to the families of UPOs present in the system. These geometrical structures are the phase space mechanisms responsible for controlling access to the potential wells, and hence selectivity can be naturally explained in terms of their interactions. To finish, in Section \ref{CONCL} we summarize our findings.

\section{The Hamiltonian Model}
\label{sec:model}
 
In this work we use a simplified version of the potential energy surface (PES) given in \cite{collins2013}, by considering that the potential is symmetric with respect to the $x$-axis. The reason for this assumption is that, in a setup where we know that the expected product ratio from the branching of trajectories is $1:1$, we would like to understand the underlying phase space mechanism that controls selectivity in this class of energy landscapes.

\begin{figure}[htbp]
	\begin{center}
		\includegraphics[scale=0.4]{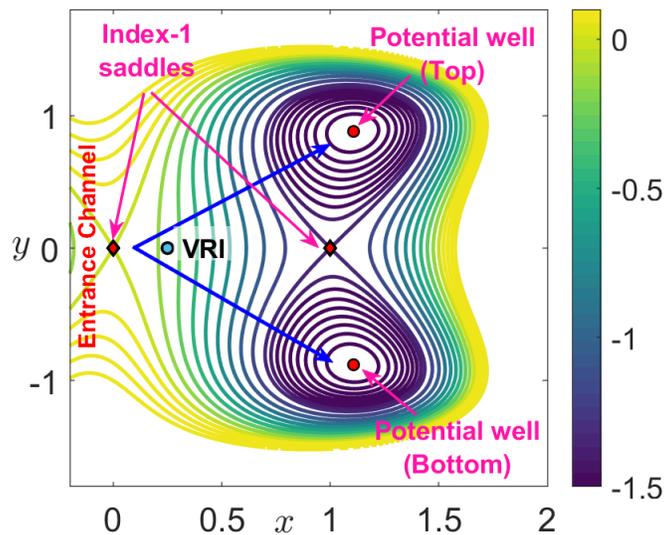}
	\end{center}
	\caption{Equipotential contours of the potential energy surface described in Eq. \eqref{pes_modelVRI}. We have marked index-1 saddles (upper and lower) as red diamonds, minima of the potential wells with red dots and the valley-ridge inflection point as a blue point. The blue arrows indicate the possible paths for trajectories that enter the system through the upper channel of the system, illustrating the chemical selectivity mechanism.}
	\label{pes_conts}
\end{figure}

The key elements of the topography of our PES are: an exit/entrance channel that is characterized by an index-1 saddle (upper index-1) and an index-1 saddle (lower index-1) which is an energy barrier separating two potential wells. Moreover the PES has a valley-ridge inflection point (VRI), which is located between both index-1 saddles. An equipotential map of the PES together with its main topographic characteristics is displayed in Fig.\ref{pes_conts}. We have also included the location of the VRI point and the blue arrows indicate the possible fates of inco\-ming trajectories that enter the system through the channel of the high energy index-1 saddle. Table \ref{tab:tab1} gives the configuration space coordinates and energies of all the critical points.

The 2 DoF Hamiltonian model that we study is the sum of kinetic plus potential energy:
\begin{equation}
H(x,y,p_x,p_y) = \dfrac{p_x^2}{2 m_x} + \dfrac{p_y^2}{2 m_y} + V(x,y) \;,
\label{hamiltonian}
\end{equation}
where the PES has the form:
\begin{equation}
V(x,y) = \dfrac{8}{3}x^3 - 4x^2 + \dfrac{1}{2} y^2 + x y^2 \left(y^2 - 2\right) \;,
\label{pes_modelVRI}
\end{equation}  
and $m_{x}$, $m_{y}$ represent the masses of the $x$ and $y$ DoF respectively. For this study we choose $m_x = m_y = 1$, and thus Hamilton's equations of motion are the following:
\begin{equation}
\begin{cases}
\dot{x} = \dfrac{\partial H}{\partial p_x} = p_x \\[.4cm]
\dot{y} = \dfrac{\partial H}{\partial p_y} = p_y  \\[.4cm]
\dot{p}_x = -\dfrac{\partial H}{\partial x} = 8 x \left(1 - x\right) + y^2\left(2 - y^2 \right) \\[.4cm]
\dot{p}_y = -\dfrac{\partial H}{\partial y} = y \left[4 x \left(1 - y^2\right) - 1\right]
\end{cases}
\;.
\label{ham_eqs}
\end{equation}
 
\begin{table}[htbp]	
	\begin{tabular}{| l | c | c | c|}
		\hline
		Critical point \hspace{1cm} & \hspace{0.6cm} Location $(x,y)$ \hspace{0.6cm} & \text{Energy} $(V)$ & \hspace{.6cm} \text{Stability} \hspace{.6cm} \\
		\hline\hline
		Index-1 Saddle (Upper) & $(0,0)$ & 0 & saddle $\times$ center \\
		\hline
		Index-1 Saddle (Lower) & $(1,0)$ & -4/3 & saddle $\times$ center \\
		\hline
		Potential Well (Top) & $(1.107146,0.879883)$ & -1.94773 & center \\
		\hline
		Potential Well (Bottom) & $(1.107146,-0.879883)$ & -1.94773 & center  \\
		\hline
		\end{tabular} 
		\caption{Location and energies of the critical points of the PES, together with their linear stability behavior when considered as equilibrium points of Hamilton's equations.} 
	\label{tab:tab1} 
\end{table}

\section{The Phase Space Transport Mechanisms Governing Selectivity}
\label{Tr_mech}

\begin{figure}[htbp]
\begin{center}
		A)\includegraphics[scale=0.24]{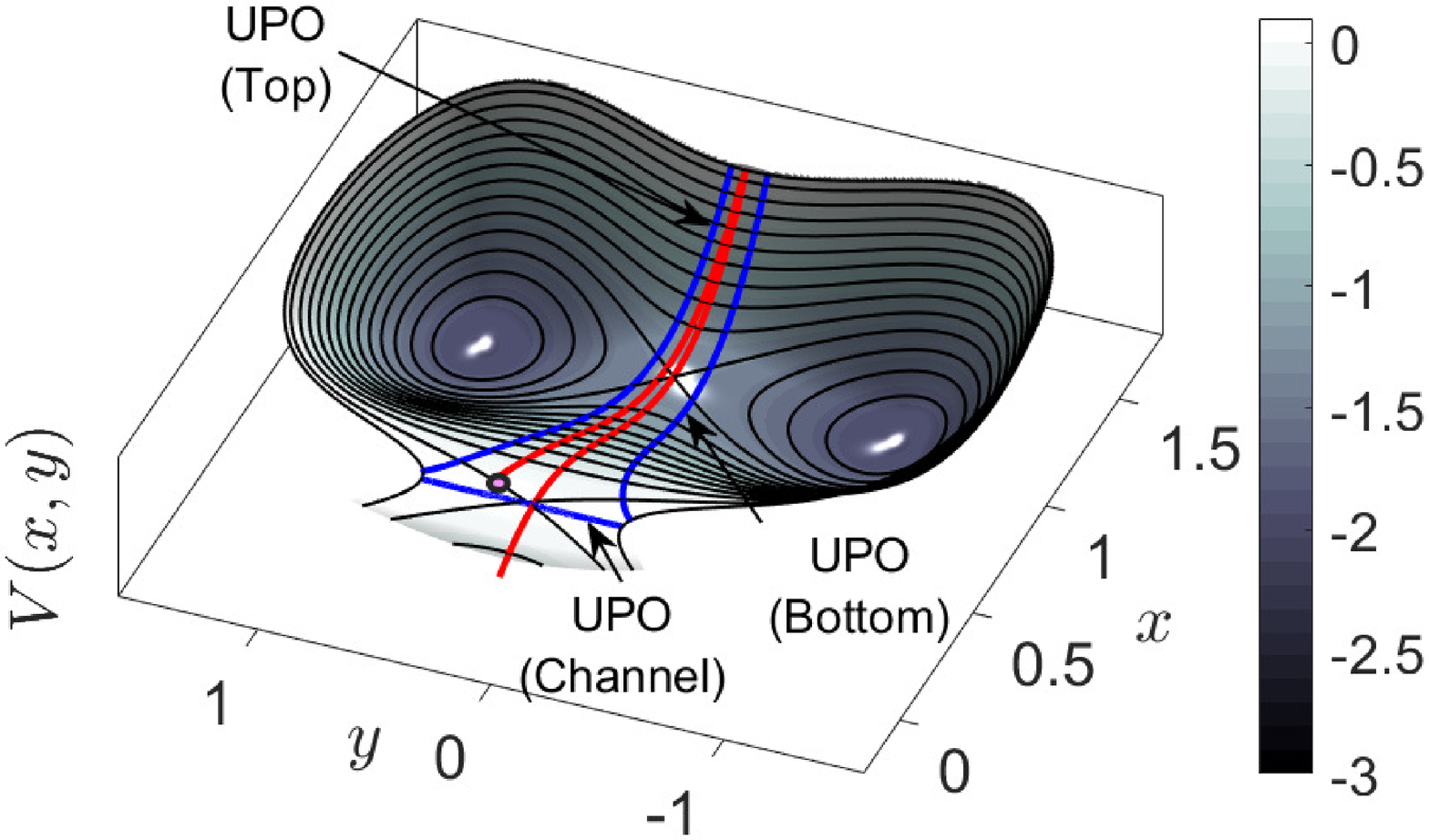}
		B)\includegraphics[scale=0.24]{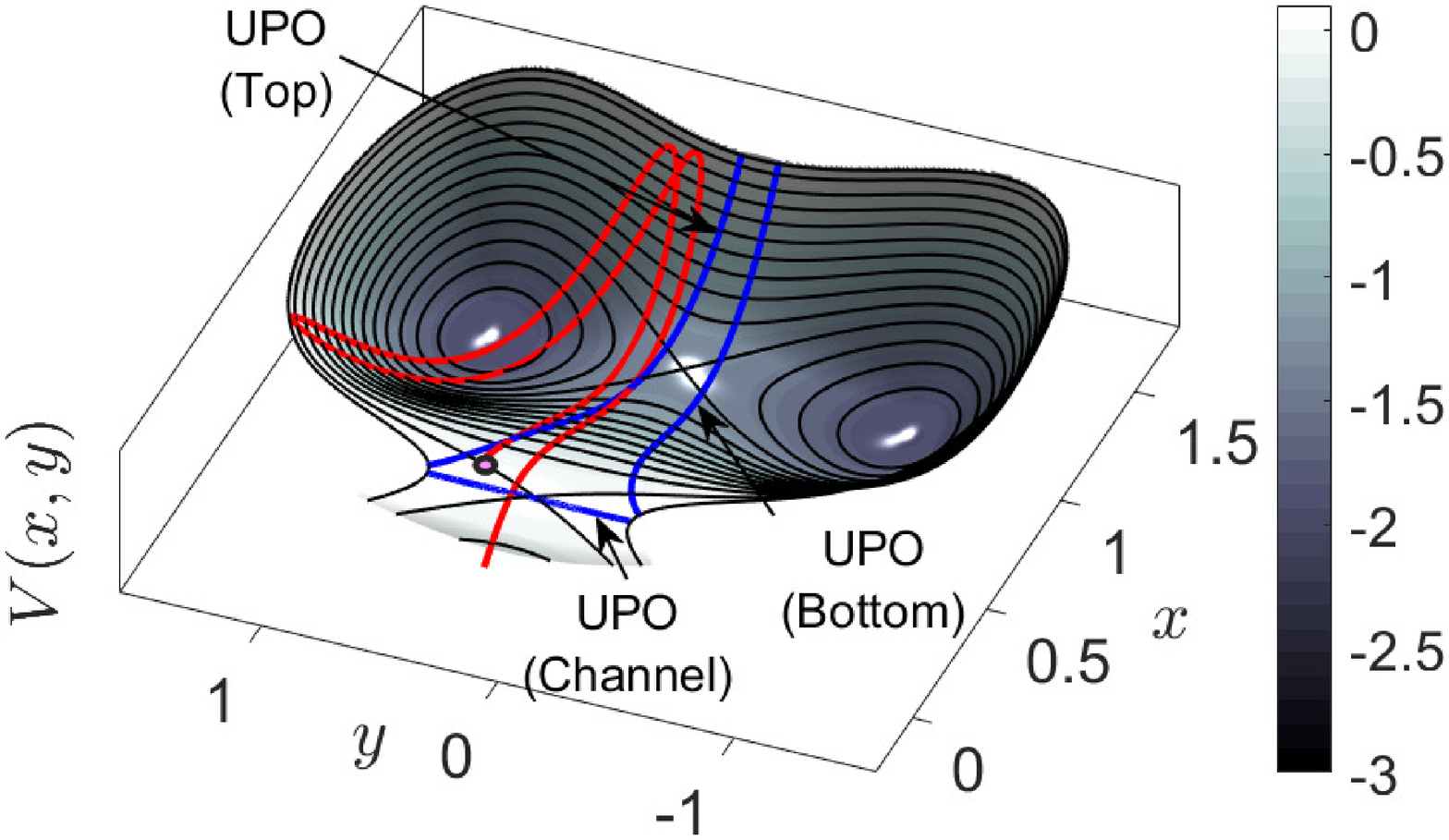}
	\end{center}
	\caption{Illustration of the different types of qualitative dynamical behavior of trajectories with energy $H_0 = 0.1$ that enter the system through the channel. We display the forward time evolution of the trajectories in red, projected onto the PES. The initial conditions are taken on the surface of section defined in Eq. \eqref{PSOS}, and are shown as magenta dots. The unstable periodic orbits of the system are depicted as blue curves.}
	\label{3d_trajs}
\end{figure}

In this section we will determine the phase space mechanisms responsible for selectivity in a way that enables us to predict the fate of individual reactive trajectories. We will work at a fixed total energy $H_0 = 0.1$. In Fig. \ref{3d_trajs} we show in red two qualitatively distinct reactive trajectories projected onto the PES. These trajectories are representative examples of the following trajectory behaviour:

\begin{enumerate}

    \item[A.] The reacting trajectory starts  from the entrance region, reaches the energy boundary without visiting either of the wells, and then returns to the exit region without visiting either of the wells.
    
    \item[B.] The reacting trajectory starts from the entrance region and visits the region of the top or bottom  well.

\end{enumerate}

Moreover, in each of the panels of  Fig. \ref{3d_trajs} we depict three curves in blue. These are the UPO of the upper index-1 saddle, the UPO in the region of the top well (top UPO) and the UPO in the region of the bottom well (bottom UPO). The UPO of the upper index-1 saddle controls  entrance and exit from the reaction region, the UPO in the region of the top well (top UPO) controls entrance and exit to the top well and the UPO in the region of the bottom well (bottom UPO) controls entrance and exit to the bottom well. This occurs through the stable and unstable manifolds of the UPOs in a way that we now explain.

For two degree-of-freedom Hamiltonian systems UPOs have two dimensional stable and unstable manifolds in the three dimensional energy surface. As a result of their dimensionality they divide the three dimensional energy surface, and since they are invariant trajectories cannot pass through these surfaces. Consequently, they are barriers to phase space transport and they therefore constrain the way in which trajectories can evolve in phase space.  To make this more explicit we focus on the notion of homoclinic and heteroclinic trajectories. A homoclinic trajectory is a trajectory that is in both the stable and unstable manifold of an unstable periodic orbit. A heteroclinic trajectory is in the stable manifold of one periodic orbit and the unstable manifold of a different periodic orbit. Homoclinic trajectories provide a mechanism for trajectories to leave the neighbourhood of the periodic orbit and to return to the neighbourhood of the periodic orbit. Heteroclinic  trajectories provide a mechanism for trajectories to leave the neighbourhood of one periodic orbit and enter the neighbourhood of the other periodic orbit. Furthermore, the existence of homoclinic and heteroclinic trajectories imply that the stable and unstable manifolds of the periodic orbits intersect in a manner  that form two dimensional regions, called lobes, whose boundaries are segments of the stable and unstable manifolds \cite{rom1990transport,wiggins1990geometry}. Since the boundaries of these lobes are invariant, the lobes ``trap’’ regions of phase space that evolve in the manner described above, i.e. a ``homoclinic lobe’’ moves toward the periodic orbit in both forward and backward time, and a ``heteroclinic lobe'' moves toward one periodic orbit in forward time and the other periodic orbit in backward time. Since unstable periodic orbits control access to different regions in the phase space an understanding of the dynamics of the homoclinic and heteroclinic lobes tells which trajectories approach the different regions.

In order to realize this phase space framework for analysing selectivity with need to compute the stable and unstable manifolds of the three UPOs and determine how their intersections govern the qualitatively distinct trajectories A,  and B. We will use the recently developed technique of Lagrangian descriptors \cite{madrid2009,mancho2013lagrangian,lopesino2017, GG2020c}. This technique was originally developed in the context of fluids mechanics, but in recent years it has been used in a growing list of applications in chemical reaction dynamics, see, e.g. \cite{ patra2018detecting, feldmaier2017obtaining, junginger2017chemical, junginger2016transition, craven2015lagrangian, craven2017lagrangian, craven2016deconstructing}. The method has  been shown to be successful in revealing, and visualizing, high dimensional phase space structures in \cite{ demian2017, naik2019a, naik2019b}.

We begin by choosing a Poincar\'e section having fixed total energy at the entrance channel defined as:
\begin{equation}
\Sigma(H_0) = \left\{ \left(x,y,p_x,p_y\right) \in \mathbb{R}^4 \;\Big| \; x = 0.05 \; , \; p_{x}\left(x,y,p_x;H_0\right) > 0 \right\}. 
\label{PSOS}
\end{equation}
We next compute the intersection of the stable manifolds of the UPOs  (shown in blue) with the Poincar\'e section and the intersection of the unstable manifolds (shown in red) with the Poincar\'e section. This figure illustrates one of the major advantages of the method of Lagrangian descriptors—it allows one to compute {\em all} of the stable and unstable manifolds in a single computation. 

\begin{figure}[htbp]
	\begin{center}
		\includegraphics[scale=0.24]{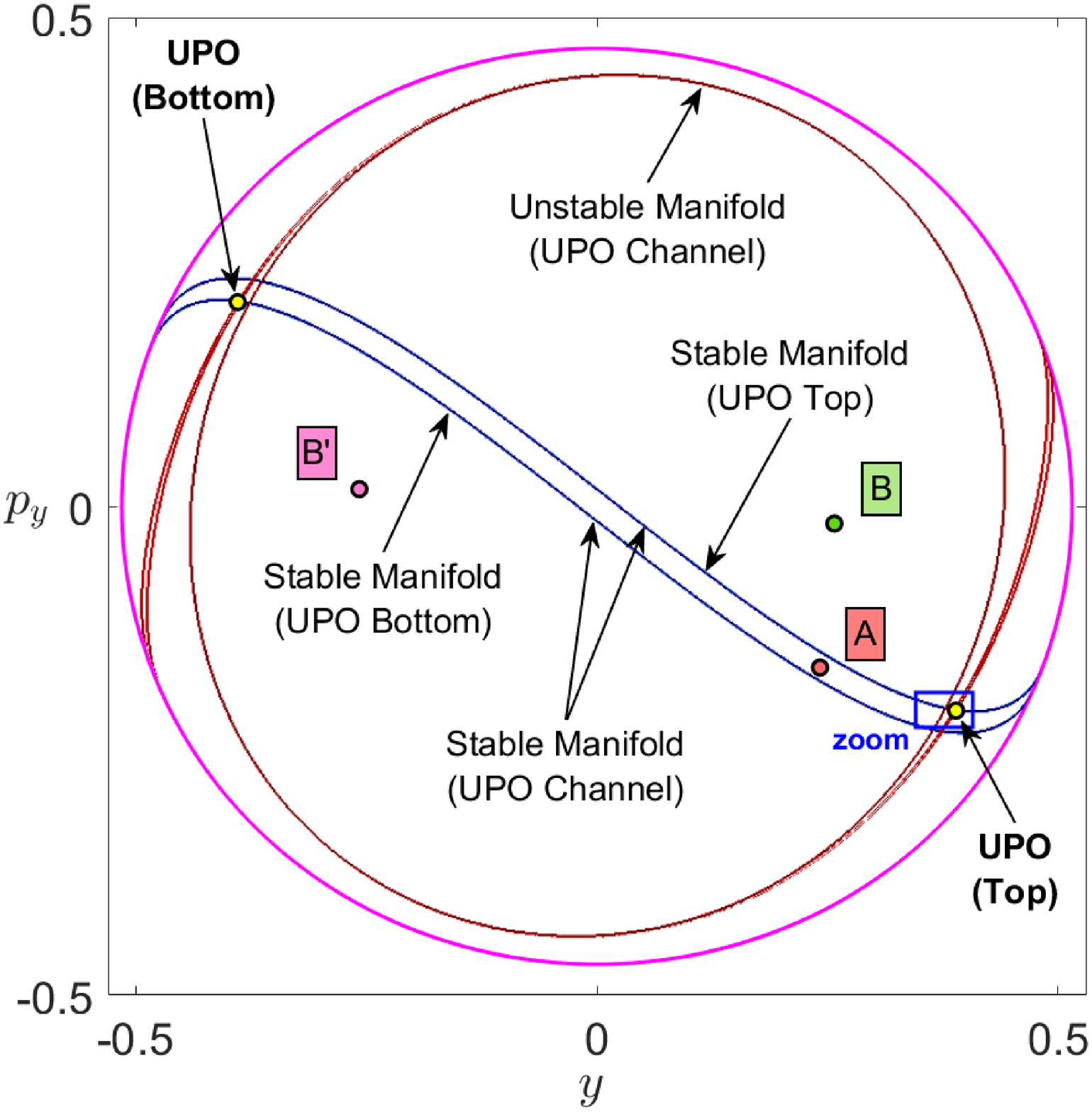}
		\includegraphics[scale=0.25]{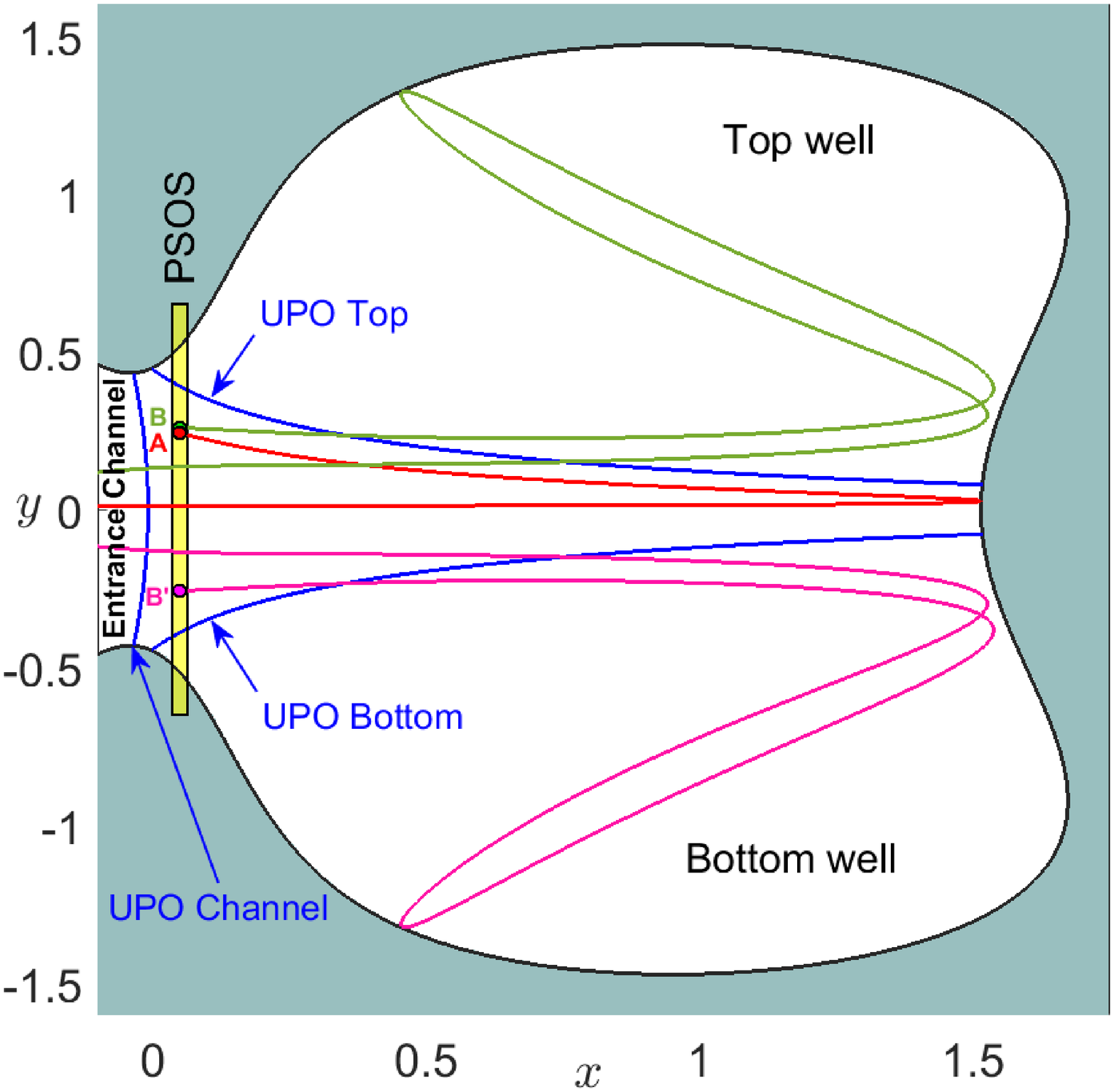} \\[.2cm]
		\includegraphics[scale=0.18]{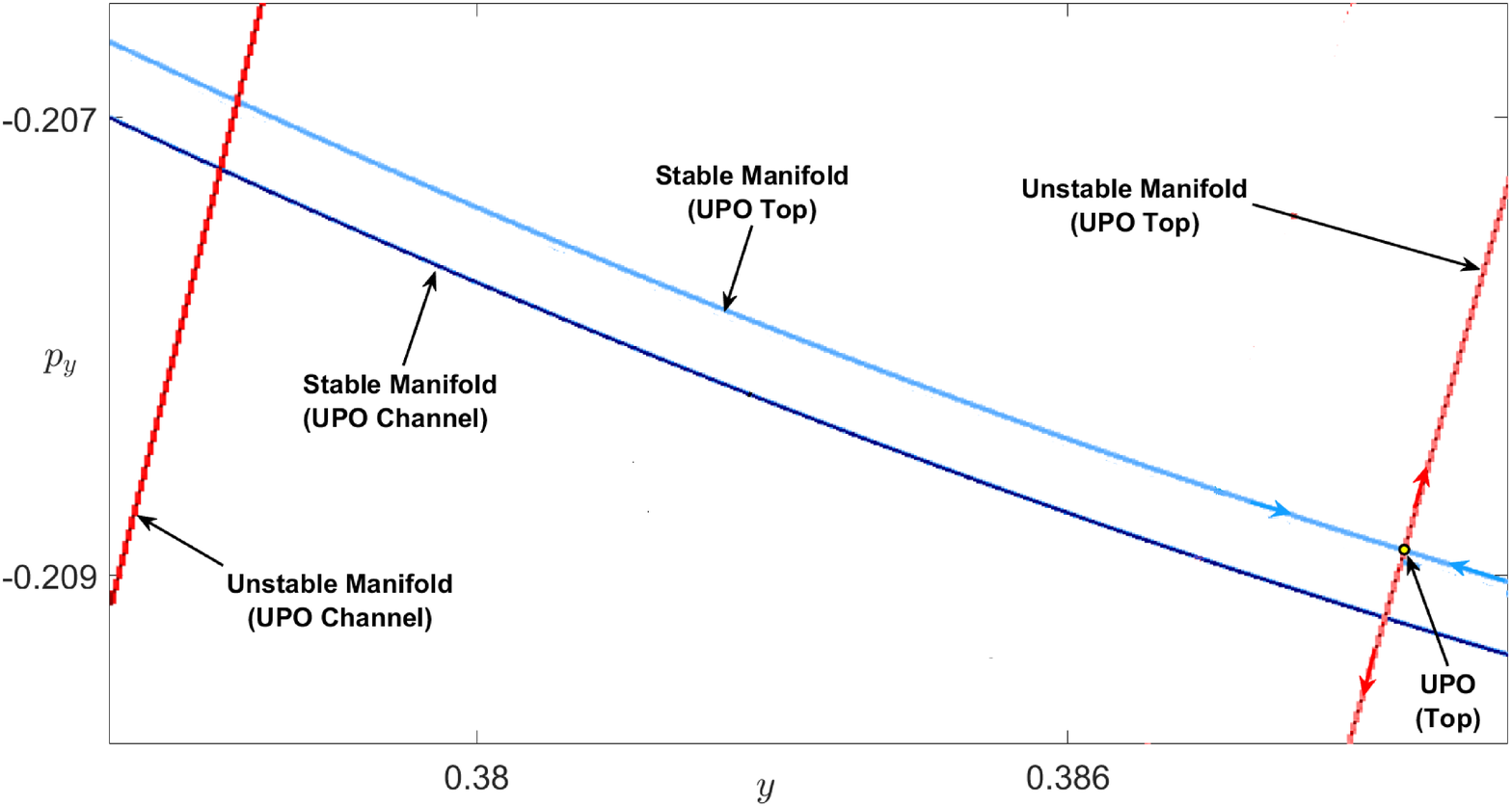}
		\includegraphics[scale=0.21]{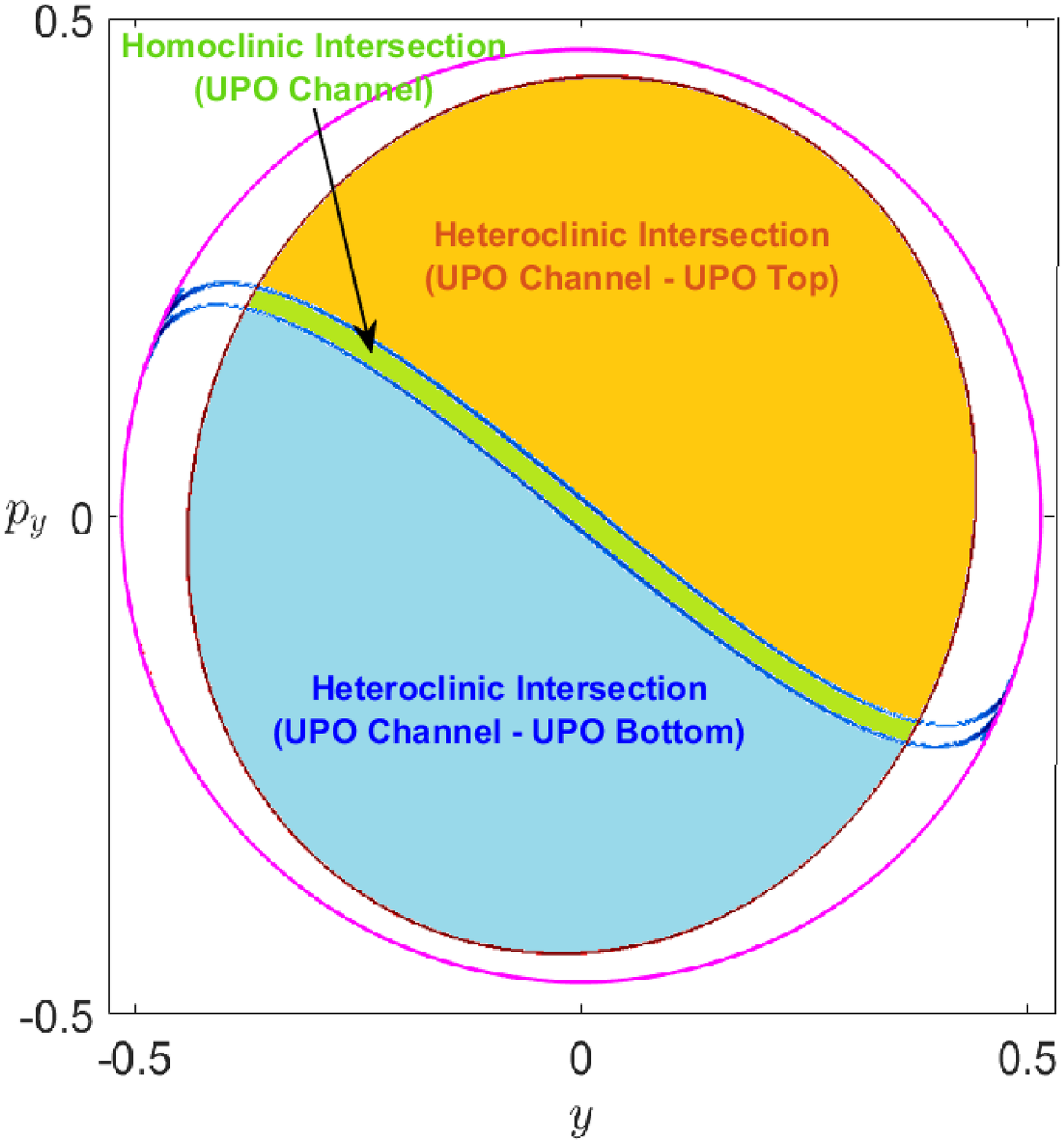}
	\end{center}
	\caption{Phase space structure and forward time evolution of trajectories with energy $H = 0.1$ used for the qualitative analysis of the system dynamics. On the top row, the left panel shows the invariant stable (blue) and unstable (red) manifolds extracted from the computation of Lagrangian descriptors on the Poincar\'e surface of section (PSOS) described in Eq. \eqref{PSOS}.  The arrows indicate the unstable manifold of the UPO of the channel and the stable manifolds of the UPO top, UPO bottom and UPO channel. Please note that every line splits into two lines (i.e. the stable manifold of the UPO channel and the stable manifold of the UPO top as we can see in the left figure on the bottom row). We also indicate the location of the UPOs responsible for controlling the access of trajectories to the top and bottom well regions. Three different initial conditions, labeled A, B and B' are selected in order to illustrate the selectivity mechanism. On the right panel, we depict the evolution of the corresponding trajectories projected onto configuration space. On the bottom row, the left image shows a zoom of the phase space region in the neighborhood of the top UPO in order to reveal its manifolds, and the right panel displays the heteroclinic and homoclinic lobes governing the selectivity mechanism.}
	\label{lobe_dyn}
\end{figure}

The initial conditions A and B in Fig. \ref{lobe_dyn} correspond to the trajectories in the panels A  and B of Fig. \ref{3d_trajs}. These initial conditions are representative examples of two  mechanisms of transport in  systems with two wells and an entrance channel. We will now explain these mechanisms in terms of the stable and unstable manifolds of the UPOs.

\begin{itemize}
\item {\bf The first mechanism - for A:} This mechanism is responsible for the transport of  trajectories into and out of the reaction region and it is mediated by the intersection of the unstable manifold of the UPO associated with the upper index-1 saddle with its stable manifold (see the trajectory that corresponds to the initial condition A in Fig. \ref{lobe_dyn} that is in the green lobe in the lower right panel of Fig. \ref{lobe_dyn}). 
\item {\bf The second mechanism - for B}: The second mechanism involves the  transport from the entrance region to the region of the   top or bottom well and is mediated by the intersection of the unstable manifold of the UPO associated with the upper index-1 saddle with the stable manifold of the top UPO (an example is the trajectory with the initial condition B in Fig. \ref{lobe_dyn} that is in the yellow lobe in the  lower right panel of Fig. \ref{lobe_dyn}), or bottom UPO (an example is the trajectory with the initial condition B$^{'}$ in Fig. \ref{lobe_dyn} that is in the blue lobe in the lower right panel of Fig. \ref{lobe_dyn}) controlling access to the top  or bottom well, respectively.  

\end{itemize}

It is accepted that due to symmetry of the PES that branching ratio of reacting trajectories is 1:1. Our work provides a dynamical justification of this result. More precisely, the symmetric transport of trajectories is reflected by the symmetry ($180^{o}$ with respect to the point (0,0)) of the invariant manifolds of the unstable periodic orbits  that are responsible for this transport (as we can see in the upper left panel of  Fig. \ref{lobe_dyn}). This means that there is a symmetry ($180^{o}$ with respect to the point (0,0)) between the lobes that are associated with the heteroclinic intersections which correspond to the transport from the entrance region to the region of the top wells (as the yellow lobe in the lower right panel of Fig. \ref{lobe_dyn})    and  the lobes that are associated with the heteroclinic intersections which correspond to the transport from the entrance region to the region of the bottom wells (as the blue lobe in the lower right panel of Fig. \ref{lobe_dyn}). Consequently, the initial conditions B and B$^{'}$ (in the upper left panel of  Fig. \ref{lobe_dyn})  have this symmetry and B corresponds to the transport from the entrance region to the region of the top well and B$^{'}$ corresponds to the transport from the entrance region to the region of the bottom well.

\section{Summary and Outlook}
\label{CONCL}

In this letter we have provided a dynamical explanation for the selectivity mechanism, which is of relevance for many organic chemical reactions. Our phase space analysis reveals that this phenomenon is governed by the heteroclinic and homoclinic interactions between the invariant manifolds associated with the UPOS that control reaction and entrance to the two wells. More precisely, these phase space geometrical structures provide us with the underlying phase space transport mechanism responsible for guiding the trajectories that enter the system through the phase space bottleneck defined in the neighborhood of the upper index-1, towards any of the wells and vice versa, therefore providing us with the theoretical and computational framework to quantify the branching ratio of  chemical reactions.

\section*{Acknowledgements}
The authors acknowledge the support of EPSRC Grant No. EP/P021123/1 and Office of Naval Research Grant No. N00014-01-1-0769.

\bibliography{main}

\end{document}